\documentclass{article}
\usepackage[a4paper, total={16cm,25cm}]{geometry}
\usepackage[utf8]{inputenc}
\usepackage{nameref,hyperref}
\usepackage[aboveskip=10pt,labelfont=bf,labelsep=period,singlelinecheck=off,font=small]{caption}
\usepackage{graphicx}
\usepackage{epstopdf}
\usepackage{authblk}
\usepackage{titlesec}
\titlelabel{\thetitle.\;}
\usepackage{indentfirst}
\usepackage{amsfonts}
\usepackage{amsmath}
\usepackage{upgreek}
\renewcommand\vec[1]{%
  \ifcat\noexpand#1\relax   
    \boldsymbol{\uprho}     
  \else
    \mathbf{#1}
  \fi
}
\usepackage{dutchcal}
\def\mc{\mathcal}
\usepackage{siunitx}
\usepackage[numbers,sort&compress]{natbib}

\title{Roughness spectroscopy of particle monolayer:\\
Implications for spectral analysis of the monolayer image}
\author[1,\footnote{Corresponding author, e-mail address: \href{mailto:ncwerons@cyf-kr.edu.pl}{ncwerons@cyf-kr.edu.pl}}]{Pawe{\l} Wero{\'n}ski}
\author[1]{Karolina Pa{\l}ka}
\affil{\footnotesize Jerzy Haber Institute of Catalysis and Surface Chemistry, Polish Academy of Sciences \authorcr
Niezapominajek 8, PL-30239, Krakow, Poland}
\date{\footnotesize\today}

\begin{document}
\maketitle

\begin{abstract}
Surface roughness of particle monolayer is one of its fundamental characteristics, conventionally described in terms of power spectral density. We derive a general equation for this function and demonstrate its similarity to those applied in the theory of scattering. We validate our analytical results by comparison with numerical results for a system with parameters corresponding to typical experimental conditions. We also demonstrate a practical application of our approach for spectral analysis of the monolayer image. Our theoretical model provides a general framework for a cheap and easy parametrization of particle monolayers.
\end{abstract}


\section{Introduction}
Geometrical effects play a key role in the interface science influencing, among others, the surface roughness. This characteristic is one of the main features of interfacial boundaries; determining friction, contact mechanics, surface wearing, adhesion, sealing~\cite{Persson2005a}, surface wettability~\cite{Yang2006}, microfluidics~\cite{Cao2009a}, reflection of electromagnetic waves~\cite{Schroder2011}, structural coloration~\cite{Parker2006a}, electrochemistry~\cite[page 166]{Bard2001}, and many others. Therefore, controlling the surface roughness is a major problem in a number of applications. One of the simplest ways to form reproducible surfaces of well-defined roughness is deposition of particles on a smooth substrate. We can do it with a variety of methods such as self-assembly at the gas-liquid interface \cite{Geisel2014}, external fields \cite{Weronski2013}, or control drying \cite{Rengarajan2005}. Such particle monolayers have been used as lithography masks \cite{Geisel2014}, to enhance the solar cell performance \cite{Loiko2017}, produce photonic crystal waveguides \cite{Rengarajan2005} or catalysts \cite{Sasaki2020}.

Once we have produced the monolayer, we may need to quantify its surface roughness. For that, we can employ advanced techniques such as grazing incidence small angle X-ray or neutron scattering \cite{Muller-Buschbaum2003, Muller-Buschbaum2013}. They require, however, access to a neutron source or synchrotron facility. With the rapid development of digital technologies and common access to different types of scanning microscopy, imaging particle analysis has emerged as the principle, potentially non-invasive method that can be very useful for monolayer characterization. The currently applied methods of image analysis mostly rely, however, on the direct identification of individual particles. That can be a major challenge in case of close-packed monolayers or monolayers with clusters of tightly packed particles \cite{Li2021}. In addition, the cost of numerical calculations increases rapidly with the number of particles per frame. Therefore, in what follows we have proposed an alternative, indirect approach to the problem, based on fitting monolayer parameters to the power spectral density (PSD) of a function describing the monolayer surface. The approach is particularly attractive considering that most of software packages for digital image analysis today are equipped with the necessary procedures.

Similar statistical methods have already been applied for digital analysis of macroscopic grain size in the investigation of granular materials \cite{Buscombe2009}. With respect to thin films and nanorough surfaces, however, PSD of microscopic images has been almost exclusively used for more or less self-affine structures. A good review of this subject is given by, e.g., Mwema and collaborators \cite{Mwema2019}. Therefore, below we have just shortly discussed the most important reports. In 1999, Strizhak et al. \cite{Strizhak1999} presented multifractal analysis of thin polycrystalline gold films imaged with atomic force microscopy (AFM). PSD analysis of the images revealed the slope of log-log plot equal about minus two. For lipid bilayers used for detection of antimicrobial peptide insertion, the slope was found to be about minus three \cite{Zhang2002a}. Similar values were reported by Rossetti et al. \cite{Rossetti2008} for regenerated cellulose films and by Sharma et al. \cite{Sharma2017} for self-assembled monolayers of 3-amino propyl tri-methoxy silane. For ultra-thin PMMA films, Pandit and coworkers \cite{Pandit2014a} obtained the slope about minus four. Even larger magnitudes, from minus four to minus five, obtained Ponomareva and collaborators \cite{Ponomareva2013} for nanocrystalline Pb(Zr,Ti)O$_3$ and TiO$_2$ thin films. Karan and Mallik studied AFM images of nanostructured thin films of copper(II) phthalocyanine deposited on polished silicon surface to get the slope about minus five. For gadolinia thin films, on the other hand, Sahoo and collaborators \cite{Sahoo2006} reported the value of about minus two and a half. In 2007, Akgun and collaborators \cite{Akgun2007} investigated the PSD of AFM images of annealed polymer brushes. They got the slope value to be about minus two.

To the best of our knowledge, two papers only present experimental results for non-fractal, monodisperse particles on a smooth surface. In 1993, Fraundorf and Armbruster \cite{Fraundorf1993} imaged a layer of Poisson distributed hemispheres of radius \SI{55}{\nano\meter} on a smooth surface to calculate the azimuthally-averaged log-log roughness spectrum of the layer. Their results suggest that the corresponding PSD behaves like a squared cosine function with the upper envelope represented by a line of slope minus three. It seems that these authors coined the term "roughness spectroscopy". In the second paper, Batys et al. \cite{Batys2016} calculated the PSD of dense, monodisperse spherical microparticle layer imaged with an optical microscope. The function exhibits a number of decreasing maxima in the range of low to mid wavenumbers and is linear in the range of high frequencies, with the slope equal minus three.

Thus, the application of PSD analysis for the determination of nanoroughness has basically been limited to random, fractal-like structures usually formed by irregular surface features, where the function can be described by the inverse power law or $k$-correlation models \cite{Gong2016}. The limitation may be caused by a lack of theoretical description of PSD for monolayers of monodisperse objects. Indeed, although the theoretical models of PSD for Gaussian random surfaces \cite{Nayak1971, Nayak1973a} are widely used since the 1970s, the analytical approach for monodisperse particle random monolayers has been developed just recently \cite{Weronski2018, Weronski2021}.

In our previous papers \cite{Weronski2018, Weronski2021}, we developed a theoretical model for the PSD of random particle monolayers where the correlation decay distance is less than the system size. In view of obvious practical importance of densely packed assemblies, here we have presented a general formalism valid for any surface coverage, including partially ordered monolayers. By means of formal analysis, we have first demonstrated that equations governing the PSD of the monolayer height profile are similar to those applied in the theory of scattering. Specifically, the PSD is a linear function of squared form factor and static structure factor. We have calculated the latter for finite rectangle and disk areas. Next, to validate our analytical results, we have carried out numerical computations of discrete PSD for a system with parameters corresponding to typical experimental conditions. Specifically, we have investigated a monolayer of particles of diameter \SI{450}{\nano\meter}, formed on a disk of diameter \SI{20}{\micro\meter}, and of surface coverage $0.85$. We have also demonstrated a practical application of our approach for spectral analysis of the monolayer image. By testing the consistency of the analytical and numerical results, we have demonstrated the applicability and usefulness of the approach for quantitative analysis of particle monolayers.

\section{\label{Sec:Analytical}Analytical considerations}
Let us consider a planar area $A$ in the Cartesian coordinate system, enclosed by a curve $C_s$ in the plane $z = 0$. The area is covered by a dense, partially ordered monolayer of $N$ unit hard spheres and is otherwise smooth. We can uniquely describe the system geometry by specifying the boundary $C_s$ and set of vectors $\vec{r_j} = (x_j, y_j)$, $j = 1, ..., N$, where $x_j$ and $y_j$ are the coordinates of the $j$-th sphere center projection on the plane $z = 0$. In this paper, we have assumed that the curve $C_s$ is only imaginary, i.e., it has no effect on the monolayer structure. In practice, we can meet the requirement by a conceptual separation of our system from a much larger monolayer or, approximately, by imposing periodic boundary conditions.

The 2D height profile of particle monolayer describes the function

\begin{equation}
  z(\vec{r}, \vec{r_1}, ..., \vec{r_N}) = \sum_{j=1}^{N} z_j(\vec{r}, \vec{r_j}),
  \label{Eq:z(r)}
\end{equation}

\noindent where $\vec{r} = (x, y)$ is the position vector in the plane $z = 0$ and

\begin{equation}
  z_j(\vec{r}, \vec{r_j}) = H \left( 1 - |\vec{r} - \vec{r_j}| \right) \left[1 + \left( 1 - |\vec{r} - \vec{r_j}|^{2} \right) ^{1/2} \right].
  \label{Eq:zj(r,rj)}
\end{equation}

\noindent Here, $H(x)$ denotes the Heaviside step function.

It is easy to show \cite{Weronski2021} that the Fourier transform (FT) of the height function

\begin{equation}
  \hat{z}(\vec{q}, \vec{r_1}, ..., \vec{r_N}) = \frac{1}{2 \pi} \int_{\mathbb{R}^2} z(\vec{r}, \vec{r_1}, ..., \vec{r_N}) \exp(-i \vec{q} \cdot \vec{r}) \mathrm{d} \vec{r} = \hat{z}_0(q) \sum_{j=1}^{N} \exp(-i \vec{q} \cdot \vec{r_j}),
  \label{Eq:Z(q)}
\end{equation}

\noindent where $i$ is the imaginary unit and $\vec{q}$ is the wave vector of the modulus $q = 2 \pi / \lambda$. Here, $\lambda$ means the dimensionless wavelength. In Eq.~\eqref{Eq:Z(q)}, $\mathbb{R}^2$ represents the set of all points in the plane $z = 0$ and $\hat{z}_0(q)$ is the FT of unit sphere at the point $(0, 0, 1)$:

\begin{equation}
  \hat{z}_0(q) = \frac{J_1(q)}{q} - \frac{\cos(q)}{q^2} + \frac{\sin(q)}{q^3},
  \label{Eq:Z0(q).sphere}
\end{equation}

\noindent where $J_k(q)$ denotes the $k$th order Bessel function of the first kind.

Consequently, the PSD of sample monolayer equals \cite{Weronski2021}

\begin{equation}
  C(\vec{q}, \vec{r_1}, ..., \vec{r_N}, A) = \frac{|\hat{z}(\vec{q}, \vec{r_1}, ..., \vec{r_N})|^2}{A} = \frac{N}{A} \hat{z}_0(q)^2 S(\vec{q}, \vec{r_1}, ..., \vec{r_N}),
  \label{Eq:C(q,r1,...,rN,A)}
\end{equation}

\noindent where

\begin{equation}
  S(\vec{q}, \vec{r_1}, ..., \vec{r_N}) = 1 + \frac{1}{N} \sum_{l=1}^{N} \sum_{\substack{m=1 \\ m \neq l}}^{N} \cos[\vec{q} \cdot (\vec{r_l} - \vec{r_m})].
  \label{Eq:S.cos}
\end{equation}

Considering that the summation in Eq.~\eqref{Eq:S.cos} is carried out over all ordered particle pairs and $\sin(x)$ is an odd function, we can write the last result as

\begin{equation}
  S(\vec{q}, \vec{r_1}, ..., \vec{r_N}) = 1 + \frac{1}{N} \sum_{l=1}^{N} \sum_{\substack{m=1 \\ m \neq l}}^{N} \exp[-i \vec{q} \cdot (\vec{r_l} - \vec{r_m})].
  \label{Eq:S.exp}
\end{equation}

By analogy with the terminology used in the theory of scattering, we have called $S(\vec{q}, \vec{r_1}, ..., \vec{r_N})$ and $\hat{z}_0(q)$ the static structure factor and form factor, respectively. Then, the product

\begin{equation}
  C(\vec{q}, \vec{r_1}, ..., \vec{r_N}, A) A = I(\vec{q}, \vec{r_1}, ..., \vec{r_N}) = N \hat{z}_0(q)^2 S(\vec{q}, \vec{r_1}, ..., \vec{r_N})
  \label{Eq:I(q,r1,...,rN)}
\end{equation}

\noindent is an analog of the scattering intensity, which we have called the roughness intensity.

The ensemble average of Eq.~\eqref{Eq:S.exp} is

\begin{equation}
  S(\vec{q}, N, A) = 1 + \frac{1}{N} \int_{A} n(\vec{r}) \int_{A} n(\vec{r'}, \vec{r}) \exp[-i \vec{q} \cdot (\vec{r'} - \vec{r})] \mathrm{d} \vec{r'} \mathrm{d} \vec{r},
  \label{Eq:S(q,N,A)}
\end{equation}

\noindent where $n(\vec{r})$ and $n(\vec{r'}, \vec{r})$ are the ensemble-averaged, areal particle number densities at the position corresponding to $\vec{r}$ and at the position corresponding to $\vec{r'}$ given a particle at the position corresponding to $\vec{r}$, respectively. In what follows, we have assumed that our system is statistically homogeneous. Thus, $n(\vec{r}) = N / A \equiv \bar{n}$ and

\begin{equation}
  n(\vec{r'}, \vec{r}) = n(\vec{r'} - \vec{r}) = \bar{n} g(\vec{r'} - \vec{r}, \bar{n}),
  \label{Eq:n(r',r)}
\end{equation}

\noindent where $g(\vec{r'} - \vec{r}, \bar{n})$ is the pair-correlation function in the limit $A \to \infty$. In general, we can determine the function numerically only. Substituting this to Eq.~\eqref{Eq:S(q,N,A)} we get

\begin{equation}
  S(\vec{q}, \bar{n}, A) = 1 + \frac{\bar{n}}{A} \int_{A} \int_{A} g(\vec{r'} - \vec{r}, \bar{n}) \exp[-i \vec{q} \cdot (\vec{r'} - \vec{r})] \mathrm{d} \vec{r'} \mathrm{d} \vec{r}.
  \label{Eq:S(q,n,A)}
\end{equation}

Please note that the integrand in Eq.~\eqref{Eq:S(q,n,A)} depends on the difference $\vec{r'} - \vec{r}$ only. Therefore, inserting into the integrand the integral of the Dirac delta function $\delta[\vec{\rho} - (\vec{r'} - \vec{r})]$ over $\vec{\rho} \in \mathbb{R}^2$ and changing the integration order we get

\begin{equation}
  S(\vec{q}, \bar{n}, A) = 1 + \frac{\bar{n}}{A} \int_{\mathbb{R}^2} g(\vec{\rho}, \bar{n}) A_r(\vec{\rho}, A) \exp(-i \vec{q} \cdot \vec{\rho}) \mathrm{d} \vec{\rho}.
  \label{Eq:S.2DFT}
\end{equation}

\noindent The area

\begin{equation}
  A_r(\vec{\rho}, A) = \int_{A} \int_{A} \delta[\vec{\rho} - (\vec{r'} - \vec{r})] \mathrm{d} \vec{r'} \mathrm{d} \vec{r}
  \label{Eq:Ar.def}
\end{equation}

\noindent is a part of $A$, containing all points at $\vec{r}$ such that the corresponding points at $\vec{r'} = \vec{r} + \vec{\rho}$ also belong to $A$. Simple geometrical considerations show that $A_r$ is an overlap of two areas $A$ displaced from each other by the vector $\vec{\rho}$. If $A$ is a disk of diameter $d = 2(A / \pi)^{1/2}$, the area is independent of the direction of $\vec{\rho}$ and equals:

\begin{equation}
  A_r(\vec{\rho}, A) \equiv A_r(\rho, d) = H \left( 1 - \frac{\rho}{d} \right) \frac{d^2}{2} \left\{ \arccos \left( \frac{\rho}{d} \right) - \frac{\rho}{d} \left[1 - \left( \frac{\rho}{d} \right)^2 \right]^{1/2} \right\}
  \label{Eq:Ar.cir}
\end{equation}

\begin{figure*}
  \centering
  \includegraphics[width = \textwidth]{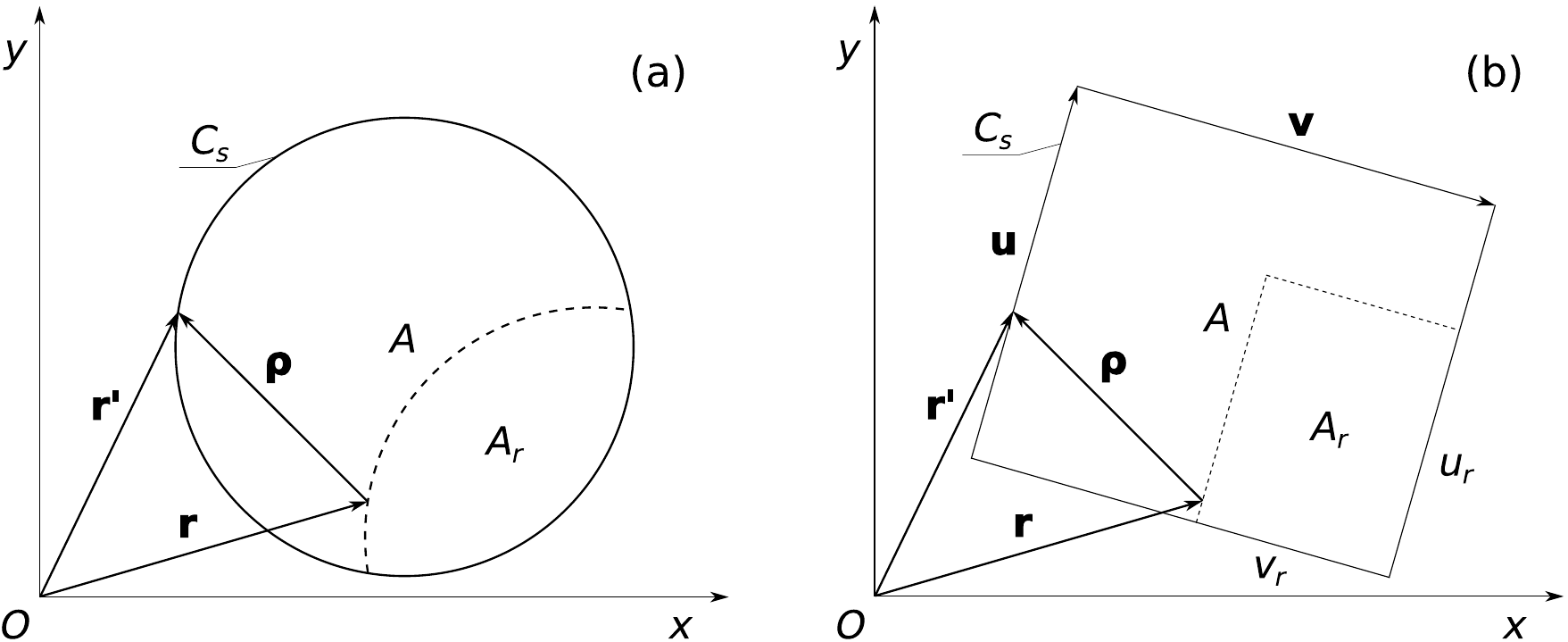}
  \caption{Surface area $A_r$ containing all points at $\vec{r}$ such that the corresponding points at $\vec{r'} = \vec{r} + \vec{\rho}$ also lie in $A$. (a) The area $A$ is a disk, Eq.~\protect\eqref{Eq:Ar.cir}. (b) The area $A$ is a rectangle, Eq.~\protect\eqref{Eq:Ar.rec}. The vector $\vec{\rho}$ is shown in two of its many extreme positions where $\vec{r'} \in C_s$.~\protect\cite{Inkscape2020}}
  \label{Fig:Ar}
\end{figure*}

\noindent [see Fig. 1(a)]. For a rectangle of the sides $\vec{u}$ and $\vec{v}$ the area equals

\begin{equation}
  A_r(\vec{\rho}, A) \equiv A_r(\vec{\rho}, \vec{u}, \vec{v}) =  u_r(\vec{\rho}, \vec{u}) v_r(\vec{\rho}, \vec{v}),
  \label{Eq:Ar.rec}
\end{equation}

\noindent where

\begin{equation}
  x_r(\vec{\rho}, \vec{x}) = H \left( 1 - \frac{\vec{\rho} \cdot \vec{x}}{x^2} \right) x \left( 1 - \frac{\vec{\rho} \cdot \vec{x}}{x^2} \right), \quad \vec{x} = \vec{u}, \vec{v}
  \label{Eq:xr(rho,x)}
\end{equation}

\noindent [see Fig. 1(b)].

With that, we can express the structure factor in the familiar form

\begin{equation}
  S(\vec{q}, \bar{n}, A) = 1 + \bar{n} \hat{g}_s(\vec{q}, \bar{n}, A),
  \label{Eq:S.gs(q,n,A)}
\end{equation}

\noindent where

\begin{equation}
  g_s(\vec{\rho}, \bar{n}, A) = g(\vec{\rho}, \bar{n}) w(\vec{\rho}, A)
  \label{Eq:gs(rho,n,A)}
\end{equation}

\noindent is the pair-correlation function in the finite area $A$. Here, $w(\vec{\rho}, A) = A_r(\vec{\rho}, A) / A$ is a weight function accounting for the boundary effect. In case of circular area $A$ we get

\begin{equation}
  w(\vec{\rho}, A) \equiv w \left( \frac{\rho}{d} \right) = \frac{2}{\pi} H \left( 1 - \frac{\rho}{d} \right) \left\{ \arccos \left( \frac{\rho}{d} \right) - \frac{\rho}{d} \left[ 1 - \left( \frac{\rho}{d} \right)^2 \right]^{1/2} \right\},
  \label{Eq:w.cir}
\end{equation}

\noindent while for the rectangular area

\begin{equation}
  w(\vec{\rho}, A) \equiv w \left( \frac{\rho_u}{u}, \frac{\rho_v}{v} \right) = H \left( 1 - \frac{\rho_u}{u} \right) H \left( 1 - \frac{\rho_v}{v} \right) \left( 1 - \frac{\rho_u}{u} \right) \left( 1 - \frac{\rho_v}{v} \right),
  \label{Eq:w.rec}
\end{equation}

\noindent where

\begin{equation}
  \rho_x = \frac{\vec{\rho} \cdot \vec{x}}{x}, \quad \vec{x} = \vec{u}, \vec{v}
  \label{Eq:rhox(rho,x)}
\end{equation}

\noindent is the length of projection of vector $\vec{\rho}$ on the vector $\vec{x}$. Please note that Eq.~\eqref{Eq:w.cir} is consistent with the results reported by Peter Krüger and collaborators \cite{Kruger2013}.

In what follows, we have assumed that our system is statistically isotropic, i.e., $g(\vec{\rho}, \bar{n}) \equiv g(\rho, \bar{n})$ and area $A$ is a disk of diameter $d$. Then, Eq.~\eqref{Eq:S.2DFT} simplifies to

\begin{equation}
  S(\vec{q}, \bar{n}, A) \equiv S(q, \bar{n}, d) = 1 + 2 \pi \bar{n} \int_2^d g(\rho, \bar{n}) w \left( \frac{\rho}{d} \right) J_0(q \rho) \rho \mathrm{d} \rho.
  \label{Eq:S(q,n,d)}
\end{equation}

Eq.~\eqref{Eq:S(q,n,d)} is valid for any number density. In case of less ordered or random monolayers, if the correlation decay distance $r_c \ll d / 2$, we can find the partially closed solution of the integral on the RHS of Eq.~\eqref{Eq:S(q,n,d)} \cite{Weronski2018}:

\begin{equation}
  S(q, \bar{n}, d) = 1 + 4 \pi \bar{n} \left\{ \left[ \frac{J_1(q d / 2)}{q} \right]^2 - \frac{J_1(2q)}{q} + \frac{1}{2} I_c(q, \bar{n}) \right\},
  \label{Eq:S(q,n,d).Ao<<A}
\end{equation}

\noindent where

\begin{equation}
  I_c(q, \bar{n}) = \int_{2}^{r_c} \left[ g(\rho, \bar{n}) - 1 \right] J_0(q \rho) \rho \mathrm{d} \rho.
  \label{Eq:Ic(q,Q)}
\end{equation}

In the limiting case of low surface coverage where, usually, the radial distribution function (RDF) can be approximated with a step function and $r_c = 2 \ll d / 2$, Eq.~\eqref{Eq:S(q,n,d).Ao<<A} reduces to the closed form

\begin{equation}
  S(q, \bar{n}, d) = 1 + 4 \pi \bar{n} \left\{ \left[ \frac{J_1(q d / 2)}{q}\right]^2 - \frac{J_1(2q)}{q} \right\}.
  \label{Eq:S(q,n,d).Q<<1}
\end{equation}

Substituting Eq.~\eqref{Eq:S(q,n,d)} to Eqs.~\eqref{Eq:C(q,r1,...,rN,A)} and \eqref{Eq:I(q,r1,...,rN)} we obtain the ensemble averaged PSD and roughness intensity of statistically homogeneous and isotropic particle monolayer:

\begin{equation}
  C(q, \bar{n}, d) = \bar{n} \hat{z}_0(q)^2 S(q, \bar{n}, d)
  \label{Eq:C(q,n,d)}
\end{equation}

\noindent and

\begin{equation}
  I(q, \bar{n}, d) = N \hat{z}_0(q)^2 S(q, \bar{n}, d),
  \label{Eq:I(q,n,d)}
\end{equation}

\noindent respectively. These equations with adequate form factors \cite{Weronski2021} are also valid for other particles of axial symmetry such as cylinders, hemispheres, and hemicapsules.

It is easy to see that Eq.~\eqref{Eq:I(q,n,d)} also describes the intensity of radiation scattering by $N$ identical atoms, within the Born approximation, if $\hat{z}_0(q)$ represents the atomic form factor. Please note, however, that our approach is different in several aspects from diffraction methods with pair distribution function (PDF) analysis. First, we have analyzed sample images instead of diffraction patterns. Consequently, we can use virtually any microscopic image without the need for synchrotron or neutron radiation. Second, we have approximated the particle by a solid sphere, neglecting its intra-molecular details. Therefore, the real space resolution in roughness spectroscopy is of the order of nanometer, not a fraction of angstrom. Also, our form factor represents the FT of height function of solid sphere, while in PDF it represents the FT of spatial density distribution of scattering object. Finally, for $\rho \geq d$ the weight function in Eq.~\eqref{Eq:S(q,n,d)} vanishes and therefore we can calculate the structure factor directly as a FT of $g_s(\rho, \bar{n}, A)$. In PDF experiments, on the contrary, the system size is macroscopic and for distances $\rho \gg 1$ the weight function is still close to unity, which makes the surface integral of RDF divergent. Therefore, to avoid technical problems with calculating the FT \cite{Shen2005}, \cite[page 133]{Loiko2018a}, the RDF in PDF methods is replaced by the total correlation function $h(\rho, \bar{n}) = g(\rho, \bar{n}) - 1$ \cite{Torquato2002}. In the scattering literature, $h(\rho, \bar{n})$ has often been called differential pair distribution function \cite{Benmore2015}.

The ensemble averaged analytical equations derived in this section provide means to characterize sample particle monolayers. Once we have determined the discrete, approximate PSD of a monolayer, we can calculate its structure factor with Eq.~\eqref{Eq:C(q,n,d)} and then, through the inverse Hankel transform of Eq.~\eqref{Eq:S(q,n,d)}, we can compute the RDF. We will validate the application of this approach to particle monolayers in our next paper, though. Here, we have used a numerically predetermined RDF to compute the structure factor and then least-squares fit Eq.~\eqref{Eq:C(q,n,d)} to numerically calculated PSD of sample monolayers generated in computer simulations. This way we have determined the particle size and monolayer number density. Comparing the predefined simulation parameters with the fitting ones we can evaluate the reliability of our method.


\begin{figure}
  \centering
  \includegraphics[width = 3.4in]{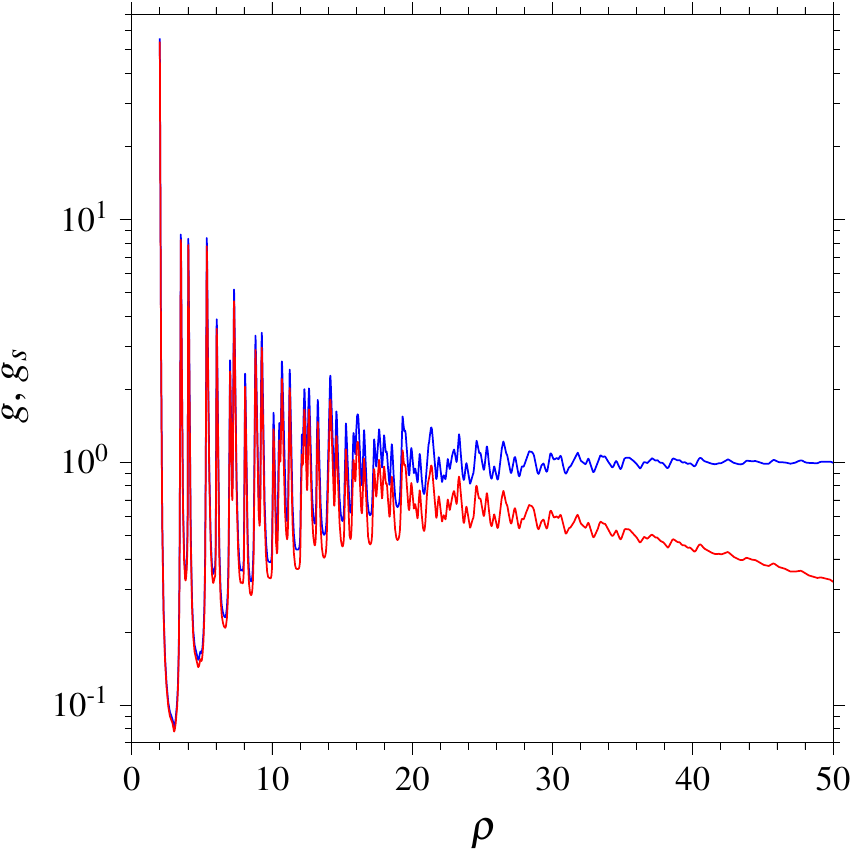}
  \caption{\label{Fig:g(rho)}RDF of partially ordered monolayer at the number density $\bar{n} = 0.27$. The blue line represents results of numerical computations for infinite area $A$, described in Sec.~\protect\ref{Sec:Numerical}. The red line denotes results for a disk area $A$ of diameter $d = 800/9$, derived from Eq.~\protect\eqref{Eq:gs(rho,n,A)}. Please note the RDFs variation at the large distance $\rho \approx 50$, characteristic for partially ordered systems.}
\end{figure}

\section{\label{Sec:Numerical}Numerical Computations}
To validate our analytical results, we have complemented them by numerical computations. They involve four types of calculations: generating the partially ordered particle assemblies, determining their RDF, evaluating the structure factor, and computing the discrete approximation of PSD. Most of the details of our computations are described in our previous paper \cite{Weronski2021} and we have not discussed them here. Instead, we focus primarily on aspects that are different.

The most basic task we have to complete before proceeding to further analysis is generating the partially ordered monolayers of hard spheres. We have produced the particle assemblies with the event-driven molecular dynamics \cite{Donev2005a,Donev2005}. Specifically, we have used the program PackLSD.64.x by Aleksandar Donev, available on the Internet \cite{PackLSD2005}. In our study, we have formed and analyzed particle systems at the surface coverage 0.85 that corresponds to $\bar{n} \approx 0.27$. We have simulated particle assemblies at square surfaces of two different sizes. To calculate the ensemble-averaged RDF and structure factor, we have generated 26 replicas of a big system with $8.5 \times 10^7$ particles, to achieve a sufficient accuracy. To calculate the discrete approximation of PSD, on the other hand, we have produced 10 replicas of a small system with 2138 particles, to approach the real experimental conditions~\cite{Weronski2021e}.

\begin{figure}
  \centering
  \includegraphics[width = 3.4in]{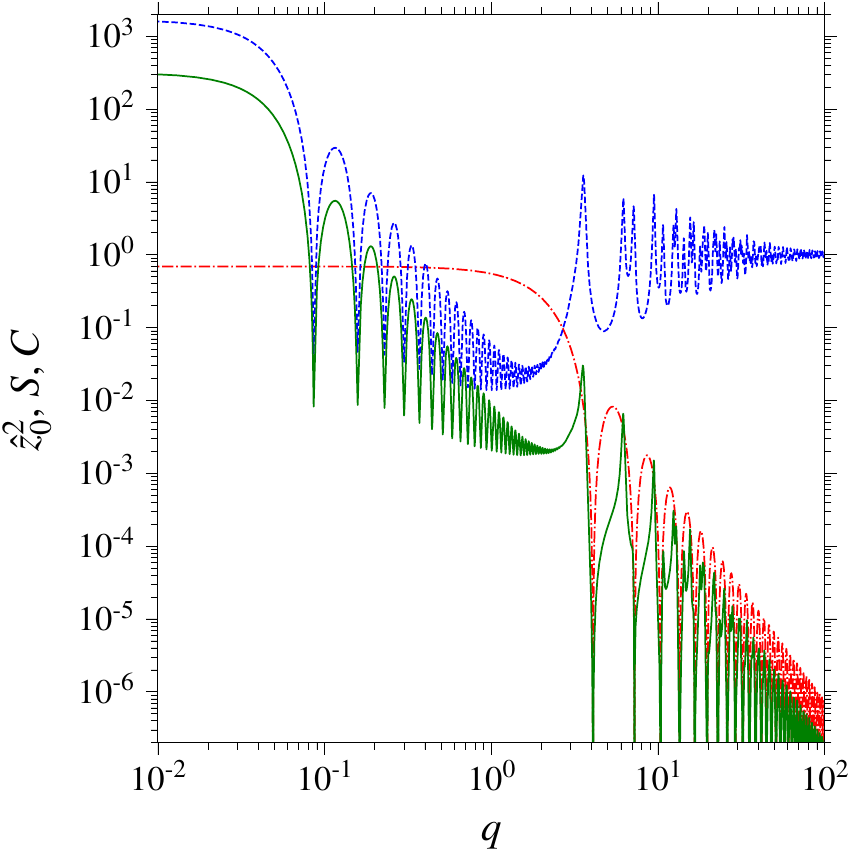}
  \caption{\label{Fig:z02,S,C(q)}Squared form factor, static structure factor, and PSD of particle monolayer calculated at the number density $\bar{n} = 0.27$ with Eqs.~\protect\eqref{Eq:Z0(q).sphere}, \protect\eqref{Eq:S(q,n,d)}, and \protect\eqref{Eq:C(q,n,d)}, respectively. Red dash-dot, blue dash, and green solid lines represent the squared form factor, structure factor, and PSD, respectively. All quantities are dimensionless, i.e., normalized by suitable powers of the particle radius.}
\end{figure}

To find a numerical approximation of RDF, we have first counted particle projections on the plane $z = 0$ within thin rings around the projection of a central particle. We have calculated the discrete approximations of RDF in the interval of $\rho$ from 2 to 90 with the ring thickness $10^{-3}$. In the narrow interval $3.9900 \leq \rho \leq 4.0020$, where the slope of RDF changes extremely rapidly, we have used the ring thickness $10^{-4}$. For each of the 26 big systems, we have determined mean values of distance and RDF in the 88108 narrow intervals, by averaging them over different central particles. Obviously, we have discarded from the averaging procedure all particles close to the boundary of the simulation area. This way, we have obtained 26 replicas of the discrete function approximating the RDF. Averaging over the replicas, we can calculate the arithmetic means of the distance and RDF in each of the 88108 intervals. We have also estimated maximum standard errors of the means and found them to be $6 \times 10^{-7}$ and $0.009$, respectively. Finally, we have computed a square B-spline representation of the discrete RDF approximation using the subroutine DFC by Richard Hanson \cite{Vandevender1982, SLATEC1993} with 3786 proper knots \cite{Weronski2021c}.

Once we have the continuous approximation of RDF, we can use Eq.~\eqref{Eq:S(q,n,d)} to compute a discrete approximation of structure factor. We have calculated its values in the wavenumber interval from $10^{-3}$ to $10^2$ at $10^5$ equidistant wavenumbers. In our computer code, we have employed the subroutine DBFQAD by Donald E. Amos, based on the highly successful integration procedure DGAUS8 developed by Rondall E. Jones \cite{SLATEC1993}. Finally, we have computed a cubic B-spline representation of the discrete approximation of structure factor. For that, we have employed the B-spline fitting procedure splrep of the package SciPy.interpolate included in the Python-based open-source library SciPy \cite{Virtanen2020}, with the knot separation distance equal about $10^{-3}$ \cite{Weronski2021d}.

To validate the equations derived in Sec.~\ref{Sec:Analytical}, we have compared their predictions with discrete PSDs computed for the replicas of small particle system, generated as described above. We have chosen dimensional parameters of the small system to mimic typical experimental conditions. Specifically, the side length of simulation area $\mc{d} = \SI{20}{\micro\meter}$ and particle diameter $2\mc{a} = \SI{450}{\nano\meter}$. Here, $\mc{d}$ denotes the dimensional diameter of circular area, inscribed in the simulation box of the area $\mc{A} = \SI{400}{\micro\squared\meter}$, which we have further analyzed. We have sampled the 2D height profile of each replica with resolution $n_r = 512$ in each direction, using a computer code. Next, using the FFTW library, ver. 3.3.9 \cite{Frigo2005, FFTW2020}, we have transformed the discrete height function, normalized the results by $n_r^2$, and calculated the power density of each wave-vector.

\section{Results and discussion}
Let us now present results of our study. To begin with, Fig.~\ref{Fig:g(rho)} presents RDFs calculated for the number density $\bar{n} = 0.27$. For a comparison, we have shown both the RDF for infinite area $A$ and for a disk area $A$ of diameter $d = 800/9$ that we have used in the simulations of our small systems. As we can see, both RDFs exhibit a number of sharp maxima similar to those observed in 2D hexagonal structures~\cite{Shao2014a}. For $\rho < 10$, the maxima are located at distances characteristic for the structures, i.e., at $\rho = 2, 2\sqrt{3}, 4, 3\sqrt{3}, 6, 4\sqrt{3},...$. At larger distances, the maxima are slightly shifted to the right and the shift increases with $\rho$. This type of behavior of RDF was reported by Batys et al.~\cite{Batys2016} for particle monolayers of coverage $0.82$. Variations of the RDFs decrease with the increase in $\rho$ but they are visible even at the distance $\rho \approx 50$. This is consistent with partial ordering of the monolayer that takes place at the high number density, close to the maximum value $\bar{n}_\infty = 0.2887$ in the crystalline state. In the infinite system, the RDF converges slowly to unity. In the finite one, the function tends to zero at the distance $\rho = d$.

Once we know the RDF, we can describe the spectral characteristics of monolayer. Figure~\ref{Fig:z02,S,C(q)} shows results of our calculations: the squared form factor, static structure factor, and PSD of the monolayer calculated with Eqs.~\eqref{Eq:Z0(q).sphere}, \eqref{Eq:S(q,n,d)}, and \eqref{Eq:C(q,n,d)}, respectively. The squared form factor is the closed form contribution to PSD and we can easily deduce its properties from Eq.~\eqref{Eq:Z0(q).sphere}. Expanding the function in power series around $q = 0$ and taking the most significant terms we get the limiting value of the function as $q \to 0$, which equals $(5/6)^2 \approx 0.7$, in agreement with the red line plot in Fig.~\ref{Fig:z02,S,C(q)}. In the opposite limit of $q \to \infty$, the first term on the RHS of Eq.~\eqref{Eq:Z0(q).sphere} becomes most significant and the squared form factor tends to $(2 / \pi) [\cos(q - 3 \pi / 4)]^2 q^{-3}$ \cite[Eq.~9.2.1]{Abramowitz1972},\cite[\href{https://dlmf.nist.gov/10.17.E3}{Eq.~10.17.3}]{Olver2019}. For the sake of clarity, we have not presented the approximation in Fig.~\ref{Fig:z02,S,C(q)}. It is easy to see, however, that the upper envelope of the function $\hat{z}_0^2(q)$ is represented by a line of slope equal minus three, just as we have expected.

\begin{figure}
  \centering
  \includegraphics[width = 3.4in]{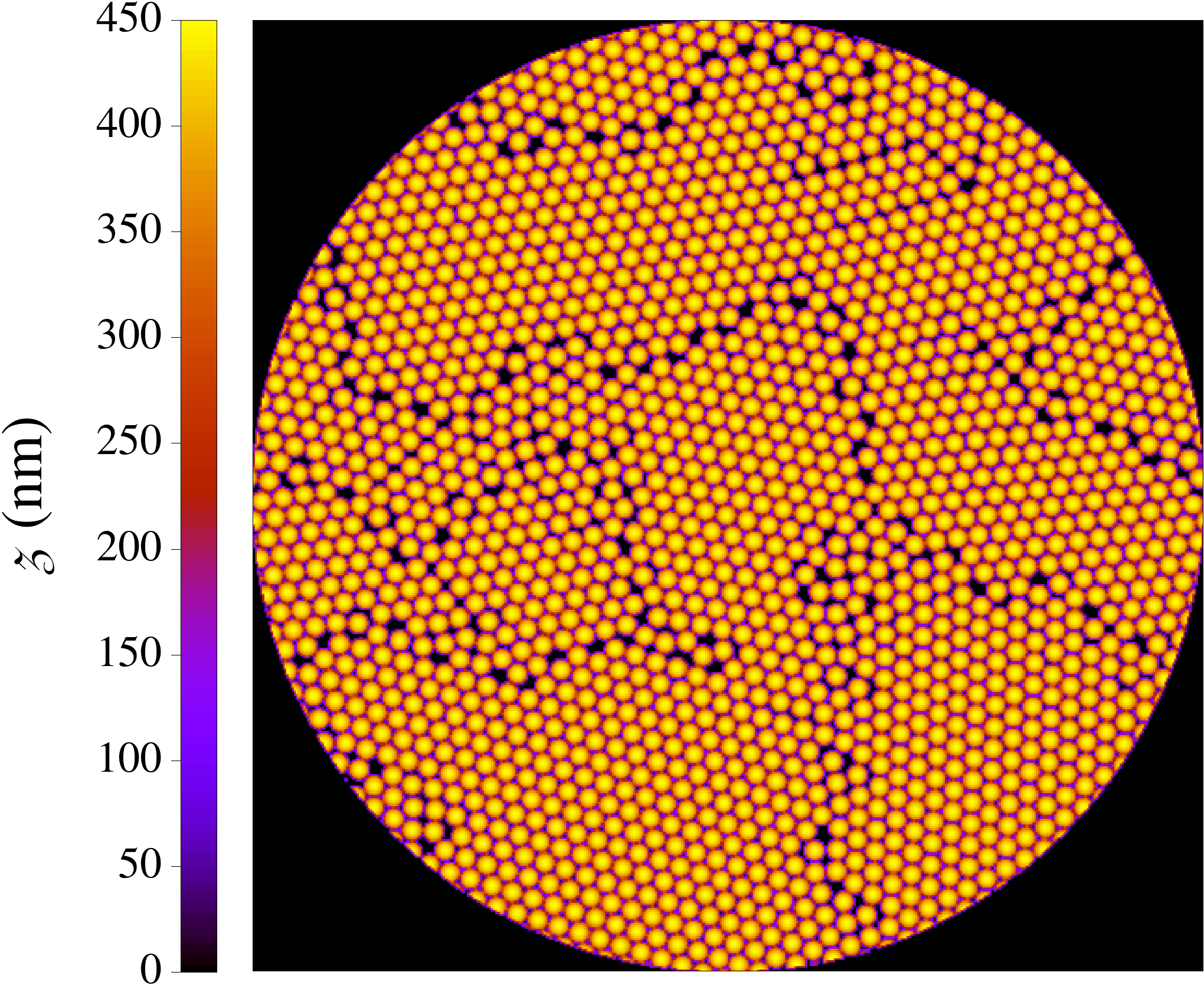}
  \caption{Image of a replica of particle monolayer at number density $\bar{n} = 0.27$, with $512 \times 512$ pixels.}
  \label{Fig:Monolayer}
\end{figure}

The limiting behavior of the static structure factor is also consistent with our theoretical predictions. To find the limiting form of the function for $q \to 0$, let's rewrite Eq.~\eqref{Eq:S(q,n,A)} in terms of total correlation function $h(\vec{r'} - \vec{r}, \bar{n}) = g(\vec{r'} - \vec{r}, \bar{n}) - 1$. That yields

\begin{equation}
  S(\vec{q}, \bar{n}, A) = 1 + \bar{n} \hat{h}_s(\vec{q}, \bar{n}, A) + \bar{n} \hat{w}(\vec{q}, A),
  \label{Eq:S.hs(q,n,A)}
\end{equation}

\noindent where $h_s(\vec{\rho}, \bar{n}, A) = h(\vec{\rho}, \bar{n}) w(\vec{\rho}, A)$ is the total correlation function on the finite area $A$. In case of disk area \cite{Weronski2018}

\begin{equation}
  \hat{w}(\vec{q}, A) \equiv \hat{w}(q, d) = \frac{1}{A} \int_{A} \int_{A} \cos[\vec{q} \cdot (\vec{r'} - \vec{r})] \mathrm{d} \vec{r'} \mathrm{d} \vec{r} = 4 \pi \left[\frac{J_1(q d / 2)}{q} \right]^2.
  \label{Eq:T0(q,n,A)}
\end{equation}

When the distance $\rho$ is large enough, the total correlation function tends to zero and so does its FT in the corresponding limit of $q \to 0$. The function $\hat{w}(q, d)$, on the other hand, tends to $\pi d^2 / 4$ \cite[Eq.~9.1.17]{Abramowitz1972},\cite[\href{https://dlmf.nist.gov/10.2.E2}{Eq.~10.2.2}]{Olver2019}. Therefore, the structure factor in this limit tends to $1 + \pi \bar{n} d^2 / 4 \approx 1680$, which we can see in Fig.~\ref{Fig:z02,S,C(q)}. In the opposite limit of $q \to \infty$, the function $J_0(q \rho)$ in the integrand on the RHS of Eq.~\eqref{Eq:S(q,n,d)} tends to $(2 / \pi)^{1/2} \cos(q - \pi / 4) q^{-1/2}$ \cite[Eq.~9.2.1]{Abramowitz1972},\cite[\href{https://dlmf.nist.gov/10.17.E3}{Eq.~10.17.3}]{Olver2019}. Then, the integral tends to zero and the structure factor tends to unity, as we can observe in the blue line plot in Fig.~\ref{Fig:z02,S,C(q)}.

It is also quite clear that the PSD changes according to the product of squared form factor and structure factor. In particular, in the limit of large wavenumber the structure factor tends to unity and therefore the upper envelope of PSD in this limit approaches a line parallel to the upper envelope of squared form factor, i.e., a line of slope equal minus three. This value is close to the experimental results reported for several self-affine structures \cite{Zhang2002a, Rossetti2008, Sharma2017} and for a dense, monodisperse spherical microparticle layer \cite{Batys2016}. The linear regime in the high frequency range of PSD shown in the later paper was, however, an artifact resulting from too wide averaging intervals, as discussed in Ref. \cite{Weronski2021}. Indeed, according to Eq.~\eqref{Eq:C(q,n,d)}, the PSD of monodisperse particles in this limit changes like $\hat{z}_0^2(q)$. Thus, in case of non-fractal objects such as spherical particles, there is no reason for linearity of PSD. Instead, it behaves like a decaying squared cosine function. We can observe this type of variation in the PSD determined experimentally by Fraundorf and Armbruster for a layer of hemispheres \cite{Fraundorf1993}. The upper envelope of their PSD is given by a line of slope minus three, in agreement with our theoretical predictions.

\begin{figure}
  \centering
  \includegraphics[width = 3.4in]{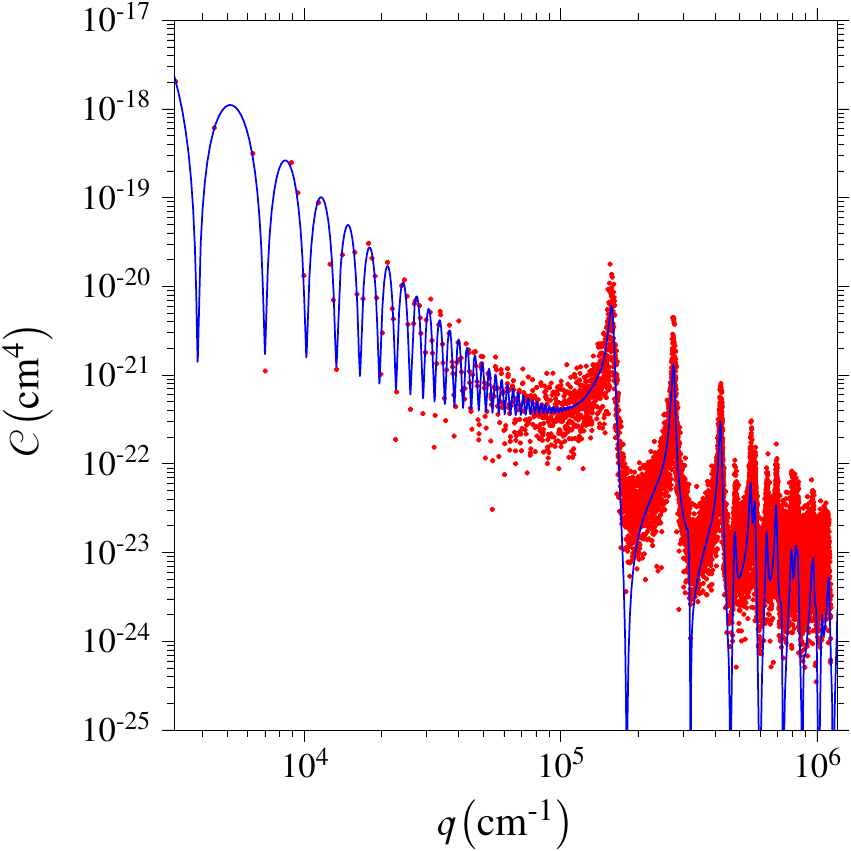}
  \caption{Dimensional PSD of the replica shown in Figure~\protect\ref{Fig:Monolayer}. Points denote the corresponding replica of discrete function $\mc{C}(\mc{q})$ calculated as described in Sec.~\protect\ref{Sec:Numerical}. The line presents predictions of Eq.~\protect\eqref{Eq:C(q,n,d,a)} with parameters $\mc{a} = \SI{225.358}{\nano\meter}$ and $\bar{n} = 0.26849$ computed by least-squares fitting. The corresponding side length of the square simulation area equals $\mc{d} = 800 \mc{a} / 9 = \SI{20.0342}{\micro\meter}$.}
  \label{Fig:C(q).fitting}
\end{figure}

Let us now discuss results of numerical validation of the analytical equations derived in Sec.~\ref{Sec:Analytical}. For that, we have first computed numerically the replicas of discrete PSD in our small system, as described in Sec.~\ref{Sec:Numerical}. Next, we have fit the dimensional version of Eq.~\eqref{Eq:C(q,n,d)} to ten replicas of the numerical data. Specifically, the dimensional PSD on the scanned, square area $\mathcal{A} = \mathcal{d}^2$ equals

\begin{equation}
  \mc{C}(\mc{q}, N, \mc{d}, \mc{a}) = \frac{N}{\mc{d}^2} \hat{\mc{z}}_0(\mc{q}, \mc{a})^2 \mc{S}(\mc{q}, N, \mc{d}, \mc{a}),
  \label{Eq:C(q,N,d,a)}
\end{equation}

\noindent where the calligraphic font denotes the dimensional quantities $\mc{q} = q / \mc{a}$, $\mc{d} = \mc{a} d$, and $\hat{\mc{z}}_0(\mc{q}, \mc{a}) = \mc{a}^3 \hat{z}_0(\mc{qa})$; whereas $\mc{S}(\mc{q}, N, \mc{d}, \mc{a})$ is the structure factor on the square area $\mc{A}$, expressed as a function of the dimensional variables. Considering that the particle number density outside the disk area is equal zero, it is easy to show that

\begin{equation}
  \mc{S}(\mc{q}, N, \mc{d}, \mc{a}) = 1 + \frac{8N}{\mc{d}^2} \int_{2\mc{a}}^{\mc{d}} g \left( \frac{\varrho}{\mc{a}}, \frac{4N \mc{a}^2}{\pi \mc{d}^2} \right) w \left( \frac{\varrho}{\mc{d}} \right) J_0(\mc{q} \varrho) \varrho \mathrm{d} \varrho = S \left( \mc{qa}, \frac{4N \mc{a}^2}{\pi \mc{d}^2}, \frac{\mc{d}}{\mc{a}} \right),
  \label{Eq:S(q,N,d,a)}
\end{equation}

\noindent where $\varrho = \mc{a} \rho$. In our computations, we have used the predetermined, ensemble averaged structure factor for the fixed $\bar{n}$ and $d$, so we can express the fitting equation as

\begin{equation}
  \mc{C}(\mc{q}, \bar{n}, \mc{d}, \mc{a}) = \frac{\pi}{4} \mc{a}^4 \bar{n} \hat{z}_0(\mc{qa})^2 S \left. (\mc{qa}, \bar{n}, d) \right|_{\bar{n}, d}.
  \label{Eq:C(q,n,d,a)}
\end{equation}

To find the parameters, we have applied $k$ the non-linear least-squares fitting procedure curve\_fit of the package SciPy.optimize \cite{Virtanen2020} with the default Levenberg–Marquardt algorithm, without rejecting any data. Figures~\ref{Fig:Monolayer} and \ref{Fig:C(q).fitting} present a replica of monolayer and corresponding PSD plot with parameters $a$ and $\bar{n}$ determined by the fitting procedure, respectively. As we can see, the agreement between discrete data and analytical prediction is good. Once we know $\mc{a}$, we can also calculate $\mc{d}$, from the fixed value of $d = 800/9$.

To quantitatively evaluate the reliability of the fitting, we have calculated the arithmetic means and standard deviations of the computed parameters. We have got the mean values of particle radius and surface coverage equal $\SI{225.11}{\nano\meter}$ and $0.853$, respectively, with the standard deviations $\SI{0.63}{\nano\meter}$ and $0.013$. Relative deviations of the means from the predefined simulation parameters are below $0.05\%$ and $0.4\%$, respectively. The relative standard deviations are below $0.3\%$ and $1.6\%$, respectively. The small values of relative deviations suggest that our method provides stable, reliable fitting parameters and can be competitive with current parametrization techniques. Indeed, to achieve similar accuracy with image processing based on individual-particle edge detection \cite{Russ1990, Sonka1993, Milne2010} or that exploiting the spherical particle symmetry \cite{Francis2006, Couteau2011, McDonald2012}, human supervision and manual corrections are necessary \cite{Weronski2013}. Without them, the error in particle number for dense particle layers is on the order of several percent. The use of machine learning approaches, on the other hand, requires big computing power and huge data sets for training. Still, the overall particle extraction rate is about 92\% and the processing time is slow \cite{Li2021}.

\begin{figure}
  \centering
  \includegraphics[width = 3.4in]{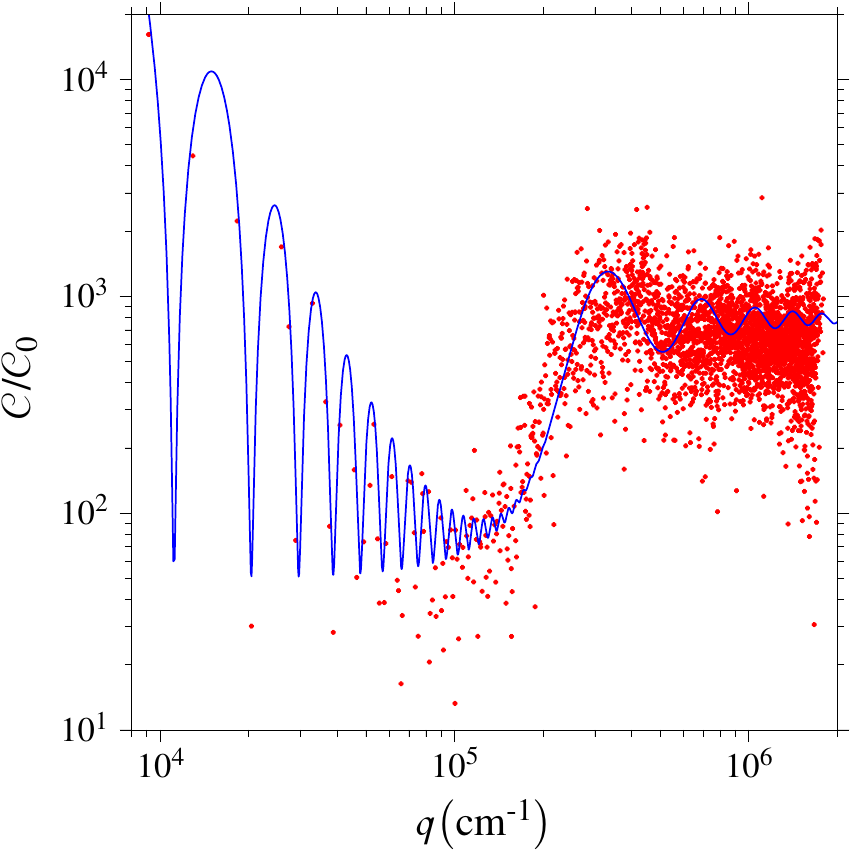}
  \caption{Ratio of PSD determined for the monolayer to average PSD of single particle. Points and lines denote experimental and theoretical results, respectively, derived for $N = 788$ particles of radius $\mc{a} = \SI{86.7}{\nano\meter}$, adsorbed on the circular area of diameter $\mc{d} = \SI{6.88}{\micro\meter}$. The surface coverage equals $0.5$. For the experimental and theoretical results, we have used the FFTW library, as described in Sec.~\protect\ref{Sec:Numerical}, and Eq.~\protect\eqref{Eq:C/C0} with Eq.~\protect\eqref{Eq:S(q,n,d).Ao<<A}, respectively.}
  \label{Fig:PSD.ratio}
\end{figure}

Although the aim of our paper is to present a proof of concept rather than a detailed study with results directly applicable in industry, let us also shortly discuss the practical use of the method. A direct use of Eqs.~\eqref{Eq:C(q,N,d,a)} or \eqref{Eq:C(q,n,d,a)} with the form factor given by Eq.~\eqref{Eq:Z0(q).sphere} requires the monolayer image to represent the actual height function. Therefore, for optimal results, monolayer images should be calibrated or at least, in case of AFM imaging, corrected for the tip size, shape, and asymmetry \cite{Odin1994, Dongmo1996, Villarrubia1997, Kopycinska-Muller2006, Belikov2009}. That, however, can be a cumbersome procedure. A much more practical approach is to use transformed Eq.~\eqref{Eq:C(q,N,d,a)}:

\begin{equation}
  \frac{\mc{C}(\mc{q}, N, \mc{d}, \mc{a})}{\mc{C}_0(\mc{q}, \mc{d}, \mc{a})} = N S \left( \mc{qa}, \frac{4N \mc{a}^2}{\pi \mc{d}^2}, \frac{\mc{d}}{\mc{a}} \right),
  \label{Eq:C/C0}
\end{equation}

\noindent where $\mc{C}_0(\mc{q}, \mc{d}, \mc{a}) = \hat{\mc{z}}_0(\mc{q}, \mc{a})^2 / \mc{d}^2$ is the PSD of single particle. We can determine the function directly from the monolayer image, simply by numerical calculating the PSD of the same image with all but one particle removed. For best results, the discrete function should be calculated for a number of single particles and then averaged. Once we have discrete values of the ratio $\mc{C} / \mc{C}_0$, we can least squares fit the data using Eq.~\eqref{Eq:C/C0} with the structure factor given by Eq.~\eqref{Eq:S(q,N,d,a)}, to find the monolayer parameters. In general, that requires to determine a 2D map of RDF for appropriate ranges of particle distance and number density, for the specific particle of interest. This procedure is out of scope of our paper and will be a subject of our future research. Below, however, we have presented a comparison of experimental and theoretical results for a monolayer image published by Ahn et al.~\cite{Ahn2005}.

Specifically, we have used a circular part of the image of spherical particle monolayer on a smooth substrate, of the diameter 278 pixels, published in Fig.~1A of Ref.~\cite{Ahn2005}. For our purpose, we have made the image background completely black, to keep the brightness value of pixels representing the substrate equal to zero. The monolayer was formed of monodisperse polystyrene particles of diameter \SI{140}{\nano\meter}. The surface coverage, determined by counting, was equal $0.326$. For SEM imaging, the monolayer was then sputter coated that resulted in the increase in particle size and surface coverage. As suggested by the authors, the structure of monolayer is similar to that predicted by the model of hard-sphere random sequential adsorption~\cite{Weronski2005}. For our analysis, we need to determine the actual values of the monolayer parameters. We can calculate the dimensional particle number density to be \SI{21.2}{\per\micro\squared\meter}. We have also counted the number of particles on the disk, $N = 788$. With that, we can calculate the diameter $\mc{d} = \SI{6.88}{\micro\meter}$. Considering that the diameter is $278$ pixels long, we get the scale factor \SI{24.8}{\nano\meter} per pixel. A closer inspection of the magnified image shows that the particle diameter after sputter coating is about seven pixels, i.e., \SI{174}{\nano\meter}. That means the coating was about \SI{16.7}{\nano\meter} thick and the actual surface coverage in the image is about $0.5$.

With these parameters at hand, we can perform spectral analysis of the image. First, we have saved 16 images of single particles of the monolayer. Next, we have calculated the PSD of the complete monolayer image and 16 single-particle images. After calculating the arithmetic mean of the single-particle PSD, we have computed the discrete values of ratio $\mc{C} / \mc{C}_0$. Figure~\ref{Fig:PSD.ratio} presents the experimental results. We can also calculate the ratio from Eq.~\eqref{Eq:C/C0}. Considering that in the model of random sequential adsorption the characteristic distance of correlation decay is about five particle radii~\cite{Tarjus1991a}, we can see that $d$ is larger than $r_c$ by almost an order of magnitude. Therefore, to calculate the structure factor, we can use the approximate Eq.~\eqref{Eq:S(q,n,d).Ao<<A} with the integral $I_c(q, \bar{n})$ calculated numerically for coverage $0.5$~\cite{Weronski2021b}. The results of our theoretical predictions are also displayed in Fig.~\ref{Fig:PSD.ratio}. As we can see, the agreement between the theoretical and experimental values is good, which confirms the robustness of our approach.

Finally, let us comment on the limits of applicability of our method. In general, they depend on the type of microscopy used and its resolution. In addition, the lower limit of particle size is dictated by the assumption of spherical particle shape. As discussed in Sec.~\ref{Sec:Analytical}, this limit is on the order of nanometer. In case of AFM, however, the radius of the tip is usually on the order of \SI{10}{\nano\meter} and therefore the minimum particle size can also be on this order. The maximum particle size is limited by the maximum image depth or vertical scan range. For AFM, e.g., this limit is, typically, on the order of micrometer.

\section{Conclusion}
The PSD of spherical particle monolayer is governed by equations similar to those in the theory of scattering. Specifically, it is a linear function of the squared form factor and static structure factor. The form factor is a FT of particle height profile and the structure factor is equal to the product of particle number density and Fourier transformed pair-correlation function, increased by one. The relationship suggests that roughness measurements of particle monolayers can be considered as a type of spectroscopy, in a wider sense of the term. So far, however, it had a qualitative sense only, emphasizing the wave-like character of results, often expressed as functions of wavenumber. Our results provide a framework for a quantitative, cheap, and easy method, based on statistical analysis of the digital monolayer image, for determining the particle pair-correlation function. The approach can be an attractive alternative to standard methods such as GISANS or GISAXS. We will investigate this subject in future studies. The PSD provides also quantitative information on important parameters of the system: the particle radius, number density, and dimensions of analyzed area. Equations derived in the paper provide means to determine the parameters with relative errors on the order less than 1\%. This accuracy seems quite satisfactory considering the high particle number density of analyzed monolayers.

\bigskip
\noindent\textbf{Funding:} K.P. has been partly supported by the EU Project POWR.03.02.00-00-I004/16. P.W. acknowledges the statutory research fund of ICSC PAS.

\bigskip
\noindent\textbf{Data Availability Statement:} The data presented in this study are openly available in Mendeley Data at \href{www.dx.doi.org/10.17632/33vhjk8wnt.1}{doi:10.17632/33vhjk8wnt.1}, \href{www.dx.doi.org/10.17632/3csw4wmjnr.1}{doi:10.17632/3csw4wmjnr.1}, and \href{www.dx.doi.org/10.17632/gwd6ck5mdr.1}{doi:10.17632/gwd6ck5mdr.1}; reference numbers \cite{Weronski2021e}, \cite{Weronski2021c}, and \cite{Weronski2021d}; respectively.

\bigskip
\noindent\textbf{Acknowledgements:} This research has been supported, in part, by PLGrid Infrastructure.

\bigskip
\noindent\textbf{Conflicts of Interest:} The authors declare no conflict of interest.

\section*{Abbreviations}{
The following abbreviations are used in this manuscript:

\bigskip
\noindent 
\begin{tabular}{@{}ll}
2D & two-dimensional\\
AFM & atomic force microscopy\\
FT & Fourier transform\\
FFTW & Fastest Fourier Transform in the West (software name)\\
PDF & pair distribution function\\
PSD & power spectral density\\
RDF & radial distribution function\\
RHS & right-hand side
\end{tabular}}

\section*{CRediT authorship contribution statement}
\textbf{Pawe{\l} Wero{\'n}ski:} Conceptualization, Methodology, Formal analysis, Validation (partially), Writing - Original Draft (partially), Writing - Review \& Editing, Supervision. \textbf{Karolina Pa{\l}ka:} Validation (partially), Investigation, Resources, Data curation, Writing - Original Draft (partially), Visualization. All authors have read and agreed to the published version of the manuscript.

\bibliographystyle{unsrt}
\bibliography{library}
\end{document}